\documentstyle[sprocl]{article}

\begin{document}

\title{A HAMILTONIAN MODEL FOR MULTIPLE PRODUCTION IN~HADRON-HADRON COLLISIONS}
\author{G. CALUCCI, D. TRELEANI}

\address{Dipartimento di Fisica teorica, Universit\`a di Trieste, Strada
Costiera 11. I-34014 and I.N.F.N. Sezione di Trieste, Italy }

\maketitle\abstracts{ A Hamiltonian eikonal model for the multiple production in
high energy hadron
hadron collisions is presented and worked out in the aim of providing a simple
frame for many different observables of these processes.
The Hamiltonian formulation ensures that it has the unitarity built in by
construction and the eikonal approximation makes easier the discussion of the 
possible spacial inhomogeneities of the hadrons.
The experimental data which are examined are the inelastic cross section and the
single and double inclusive cross sections.}

\section{ Description of the model}

A Hamiltonian model for the description of the multiple production process in
hadron-hadron collision is presented with the main aim of bringing together
different observables within a unique frame. A particular attention is given to
those features of the inelastic processes that can give informations on the
proton structure. 
\par
The physical ingredients of this model are the following:
\par\noindent
A sharp distinction is made between the soft dynamics, which provides the 
binding of the
partons in the hadrons and also the final hadronization of the shaken-off 
partons and the hard dynamics that causes the parton scattering.
\par\noindent
The hard collision gives a finite transverse momentum to the partons, which
remains however small with respect to the typical longitudinal momentum.
Hard rescattering is included, but not hard branching of the partons.
\par\noindent
The discrete quantum numbers spin and colour are not taken into account.

The model describes the hadrons as sets of bound partons which, due to the
interaction, may become finally unbound and show up as jets; the hadronization
process is not described.
\par
The only detailed kinematics is the transverse one, the whole treatment lies
within the frame of the eikonal formalism.[1,2,3]
The sharp distinction between backward and forward degrees of freedom allows the
following formulation: There are operators for the bound backward and forward 
partons $a_b,a_f$ and 
operators for the unbound partons $c_b,c_f$. Both are local in ${\bf b}$, the
transverse impact parameter of the parton, and they have the standard 
commutation relations,
where every backward operator commutes with every forward operator and every $a$
commutes with every $c$; so we can write a free Hamiltonian:
\begin {equation}
{\cal H}_o=\sum_{v=f,b} \omega \int d^2
b[a^{\dagger}_v({\bf b})a_v({\bf b})+c^{\dagger}_v({\bf b})c_v({\bf b})]
\end {equation}
The interaction that we want to describe is the hard collision of two bound
partons that give rise to two unbound partons is such a way however that they keep
their property of being either backward or forward. Thus the interaction
Hamiltonian
is written as:
\begin {equation}{\cal H}_I=\lambda \int d^2 b\,h_b({\bf b})h_f({\bf b})\quad
h_v({\bf b})=c^{\dagger}_v({\bf b})a_v({\bf b})+a^{\dagger}_v({\bf b})c_v({\bf
b})\;.\end {equation}
With this choice it results
$[{\cal H}_o,{\cal H}_I]=0\;,$
but the theory is not trivial even though it has been much simplified, the 
S-matrix can 
be written in the form ${\cal S}=\exp[-i {\cal H}\tau]$ where $\tau$ is an
 interaction time.
 \par
An alternative, and more realistic, form gives
a finite size to the hard interaction and a
discretization of the transverse plane. The size $\Delta $ is related to the
cut-off in the transverse momentum that must be put in order to be allowed to
perform perturbative calculations, so the natural choice is $\Delta \approx 
1/p_{\bot}^2$; this choice lead also to the interpretation of
$\tau\approx 1/p_{\bot}\approx \sqrt {\Delta}$. The commutation relations become
 $\;[A_{v,j},A^{\dagger}_{u,i}]=\delta _{i,j} \delta _{u,v}\;,$ and so on.
 In this way we get:
\begin {equation}{\cal H}_o=\sum_{v,j} \Omega
[A^{\dagger}_{v,j}A_{v,j}+C^{\dagger}_{v,j}C_{v,j}]\end {equation}
          \begin {equation}{\cal H}_I=(g/\sqrt{\Delta})\sum_j H_{b,j}\cdot
          H_{f,j}\quad\, \quad
         H_{v,j}=C^{\dagger}_{v,j}A_{v,j}
         +A^{\dagger}_{v,j}C_{v,j}\;.\end {equation}
         The coupling constant $g$ is dimensionless, it is related to the
         previous coupling constant by $g=\lambda/\sqrt {\Delta}$.
Correspondingly the S-matrix is
\begin {equation}{\cal S}=\prod_j {\cal S}_j\qquad ,\qquad 
{\cal S}_j=\exp[-i (g/\sqrt{\Delta})\tau H_{b,j}\cdot H_{f,j}]\;,\end {equation}
 with the previous interpretation of $\tau$, we get the simpler form:
\begin {equation}{\cal S}_j=\exp[-i g H_{b,j}\cdot H_{f,j}]\;.\end {equation}
In order to apply this model one must choose a definite initial state; it will
be factorized in the same way as the S-matrix: as far as its structure in a site
$j$ is concerned there are no strong indications. A possible choice, related to
 some theoretical ideas about the nonperturbative partonic
 structure of the hadrons,[4,5] is a local coherent state, so we write
 \begin {equation}|I>=\prod _j |I>_j\quad ,\quad |I>_j=
 \exp [-(|F_b|^2 +|F_f|^2)/2] 
       \exp[F_bA^{\dagger}_b +F_f A^{\dagger}_f]\,|>_j \end {equation}
It has to be noted that the weight
 $F$ of the coherent state may vary from site to site. For simplicity
 the index $j$ will be not written out, whenever possible.
\par
It is easer to express $|I>_j$ and especially to perform the subsequent
calculations in the basis generated by the auxiliary operators $P$ and $Q$.
\begin{equation} P=(C+A)/\sqrt{2}\quad Q=(C-A)/\sqrt{2}\;.
 \end {equation}
In terms of them on gets
 \begin {equation}{\cal H}_o=\sum_{v,j} \Omega
   [P^{\dagger}_{v,j}P_{v,j}+Q^{\dagger}_{v,j}Q_{v,j}]\quad ,\quad
           H_{v,j}=P^{\dagger}_{v,j}P_{v,j}-Q^{\dagger}_{v,j}Q_{v,j}
           \end {equation}
  \vskip 1pc
 \section {Some results of the model}
\vskip 1pc
\subsection{ Inelastic cross section}
\vskip 1pc
In the discrete formulation for the states and for the S-matrix  the inelastic
cross section is now calculated. In the basis generated by the operators $P$ and $Q$
the operator ${\cal S}_j$ is diagonal, so it is easy to calculate the matrix 
element $S_j={}_j\!<I|{\cal S}_j|I>_j$, it has the expression:
\begin {equation}S_j=N^2 \sum _{k_1\cdots k_4}{1\over {k_1!k_2!k_3!k_4!}}
  (|F_b|^2 /2)^{k_1+k_2} (|F_f|^2 /2)^{k_3+k_4}\exp[-i g(k_1-k_2)(k_3-k_4)]
  \;.\end {equation}
 The indices $k_1,k_2,k_3,k_4$ refer to the quanta created 
by$ P_b^{\dagger},Q_b^{\dagger},P_f^{\dagger},Q_f^{\dagger}$ respectively. 
The normalizing factor is $N=\exp \bigl[-(|F_b|^2+|F_f|^2)/2\bigr].$
Using the representation: 
\begin {equation}\exp[-i g(k_1-k_2)(k_3-k_4)]=(2\pi )^{-1}\int dudv 
\exp \bigl[i uv +
i \alpha u(k_1-k_2) +i \beta v (k_3-k_4)\bigr]\;,\end {equation}
with $\alpha \beta =g$,
the multiple sum in the expression of $S_j$ can be transformed into an
integral. In the final result we use the definitions $T_v=|F_v|^2$ and we have:
\begin {equation}S_j=(2\pi )^{-1}\int dudv \exp [i uv] 
 \exp \bigl[-T_b (1-\cos\alpha u)-T_f (1-\cos\beta v)\bigr]\;.\end {equation}
\vskip 1pc
When the distribution functions $F_v$ do not vary strongly from site to site one
can devise a continuum limit. We
consider a relation $F\approx f \sqrt{\Delta}$; therefore, if $f$ is not
singular, in the expression of $S_j$ the terms $|F|^2$ become small, the
exponential in the integral representation can be expanded and integrated term
by term with the result: 
 \begin {equation}S_j\approx 1-|F_b|^2|F_f|^2 (1-\cos g)+(1/2)|F_b|^2|F_f|^2
 (|F_b|^2+|F_f|^2)(1-\cos  g)^2+\dots\end {equation}
 With the normalization we are using the inelastic cross section at fixed
 hadronic impact parameter ${\bf B}$ is given by 
  \begin {equation}\sigma({\bf B})=2<I|(1-\Re {\cal S})|I>-\bigl | <I|(1-{\cal S})|I>\bigr |^2
  \;.\end {equation}
  The product of the matrix elements is expressed as the exponential of
  the sum of the logarithms and the sum $\Delta\sum_j$ is finally
  converted into the integration $\int d^2b$ with the final result: 
   \begin {equation}\sigma ({\bf B})=1-\exp\bigl[-\hat \sigma \int d^2b|f_b({\bf b})|^2 
  |f_f({\bf b-B})|^2 +\hat \sigma ^2\cdots\bigr]\end {equation}
  the parameter $\hat \sigma=2\Delta (1-\cos g)$
  has the role of elementary partonic cross section.
  \par
 The form of the inelastic cross section is quite usual,it contains however the
 explicit indication of the possible corrections due to rescattering, they
 will be discussed below, where a nonuniform model of the hadron will
  be explored.
 The cross section arises from the integration over the impact parameter, 
 the result depends very much on the properties of the distribution functions 
 $f$, which appear through their squared absolute values 
$t_v({\bf b})=|f_v({\bf b})|^2$, giving the transverse density of bound partons.
 In a first discussion the distribution is taken to be completely uniform in
 $b$ by setting:
 \begin {equation}t_b({\bf b})=\rho_b \vartheta(R-|b|)\quad,\quad
 t_f({\bf b})=\rho_f \vartheta(R-|b|)\;;\end {equation}
 elementary geometrical considerations give $|B|=2R\cos\gamma/2$ with 
 $0\le\gamma\le\pi$. The exponent in the integrand is given by the 
 partial superposition of the two disks. Since the superposition area is 
 $W= R^2 (\gamma -\sin\gamma)=\pi R^2\xi\;$ 
 The expression for the inelastic cross section is:
 \begin {equation}\sigma_{in}=2\pi R^2\int_o^{\pi}d\gamma \sin \gamma 
 \Bigl[1-\exp[-\nu \xi]\Bigr]\;,\end {equation}
 with $\nu =\hat \sigma \rho_b \rho_f \pi R^2$.
 \par
 It is possible to give a simple analytical form for $\sigma_{in}$ in the two
 limiting situations of very small or very large $\nu$. In the first case one
 gets 
 \begin {equation}\sigma_{in}=\pi R^2\cdot \nu \;.\end {equation}
 In the second case we can start from
 the expression 
 \begin {equation}\sigma_{in}=2\pi R^2[2-D(\nu)]\end {equation}
 where the real function 
 $D(\nu)$ defined by eq(17) is monotonically decreasing, for large $\nu$ it 
 results $D(\nu)\approx 2 (6\nu^2 /\pi^2)^{-1/3}\cdot\Gamma (2/3)$ so that the 
 geometrical limit of black disks $4\pi R^2$ is approached.
 \vskip 1pc
 
 \subsection{ Pair and double-pair production}
 \vskip 1pc
 The production of a pair can seen either as production of a backward parton or
 of a forward parton, the Hamiltonian being fully symmetric, so we can choose,
 arbitrarily to look at the forward particles; successively we shall investigate
 how much the rescattering processes may destroy the sharp correlation
 between backward and forward scattered partons.
 We start from the computation of the inclusive production from a single site.
 Straightforward, although lengthy calculations, easier in the basis
generated by the operators $P$ and $Q$, yield the result 
\begin {equation}<X_j> =
 {}_j\!<I|{\cal S}_j^{\dagger} C_f^{\dagger}C_f{\cal S}_j|I>_j 
 =(T_f/2)\bigl[1-\exp[-T_b(1-\cos (2g)]\bigr]\;.\end {equation} 

We can now go to the continuum limit, always under the hypothesis of smooth
distributions $t_v({\bf b})$; we find in eq(20) the function
$\cos (2g)$ instead of $\cos g$, so we must define a quantity
$\kappa=(\Delta/2) [(1-\cos (2g)]$, which is related to $\hat \sigma$ in this
way:
 $\kappa=\hat \sigma\cdot [1+\cos(g)]/2$. The two constants coincides
 evidently
 for small $g$, when both are: $\hat \sigma =\kappa= g^2\Delta$, but the 
 rescattering corrections, higher powers in $g^2$, are different.
 \par
 In the smooth continuous limit we expand the exponential of eq.
 (4.3)
  and we get the usual expression:
\begin {equation}X({\bf B})=\kappa \int d^2b\, t_f({\bf b}) t_b({\bf b-B})
\end {equation} 
and the inclusive one-particle cross section is
\begin {equation}D_1=\int X({\bf B}) d^2B \end {equation}
Analogously the two-particle inclusive cross section is found to be
\begin {equation}D_2=\kappa^2\int d^2B d^2b d^2b'
t_f ({\bf b}) t_b ({\bf b-B})t_f ({\bf b}') t_b ({\bf b'-B});.\end {equation}

 A ratio of the quantities that have been now calculated which is of
phenomenological interest is $\sigma_{\rm eff}=[D_1]^2/D_2$,[6,7]
In the limit of rigid disk, using the geometrical considerations one gets
\begin {equation}\sigma_{\rm eff}={{\pi R^2}\over {1-16/(3\pi^2)}}
\approx 2.2\pi R^2\;,\end {equation}

The previously calculated expression of $\sigma_{in}$ is really the
hard part of the total inelastic cross section, where as "hard" part we intend
the contribution of all the events where at least one hard scattering happens.
If we believe that, in going on with the total energy these events become more
and more important we would like to have this term not too small with respect to the
experimental $\sigma_{in}$, which in turn appears to be sizably larger, of about
a factor 2, with respect to $\sigma_{\rm eff}$, so in this model we would like
to approach, even though not reach, the black-hadron, limit which produces 
$\sigma_{in}=4\pi R^2$. 
 
\vskip 1pc
\subsection{ Multiplicity distribution and forward-backward correlations}
\vskip 1pc
The distribution of the multiplicities of the produced pairs is calculated
by defining the projection operator over the number of free partons. Since we
have always the sharp distinction between backward and forward particles we can
choose one of the two particle to define the produced pair: to be
definite we take the forward particle as signal of the pair production.
For a fixed site $j$ the number projector may be expressed as:
\begin {equation}{\cal P}_n={1\over {n!}}:[C^{\dagger}C]^n e^{-C^{\dagger} C}:
\end {equation}
The colon indicates the normal ordering of the $C$-operators.
From this definition one can perform calculations whose qualitative results can
be stated in this form: 
when the total production is not very copious the result may be
written as the sum of two Poissonian distributions, one reflects the incoming
coherent state the other the rescattering effect 
, if however the production in a single site is
high so that the rescatterig is very important the resulting expression may
be put into the form of a Poissonian distribution times
another factor, but this further factor is not a small correction, 
it changes in essential way the shape of the distribution.
\vskip 1pc
The same qualitative result is obtained by studying the forward-backward
correlations; we can calculate both the variance
 $\Sigma _f=<X _f^2>-<X _f>^2$
 with
 $<X _f^2>=<I|{\cal S}^{\dagger} C_f^{\dagger}C_f
 C_f^{\dagger}C_f{\cal S}|I>$
 and the covariance \par\noindent $\Sigma _{f,b}=<V>-<X_f><X_b>$ with:
   $<V> =
   <I|{\cal S}^{\dagger} C_f^{\dagger}C_f C_b^{\dagger}C_b{\cal S}|I>$
   
 The actual calculations show , as expected, that the correlation 
 \begin {equation}\kappa _{f,b}={{\Sigma _{f,b}}
 \over{ [\Sigma _f \Sigma _b]^{1/2} }} \end{equation} 
   for small values of $T_v$, and so for small production, goes to 1, but when
 $T_v$ becomes very large it goes to zero.
 
\vskip 1pc
\subsection{ Non uniform hadrons}
We wish now to explore the possibility that the hadron, and its projection
 over the transverse plane, shows strong inhomogeneities in the matter density.
 This is represented by assuming that there are black spots, that cover a
 limited amount of the transverse area, while a much fainter "gray" background
 fills uniformly the rest of the hadron.
 \begin {equation}|I_b>_j=\Bigl[x\exp [- |F_b|^2 /2] 
       \exp[F_bA^{\dagger}_b]+y\exp [-|G_b|^2/2 ] 
       \exp[G_bA^{\dagger}_b]\Bigr]|> \end {equation} 
        Since the two coherent
 states are not orthogonal the normalization condition is complicated,
 however if the two thickness are very different the cross term in
 normalization condition is very small and we are left with
$|x|^2+|y|^2\approx 1\;.$
 The expression for the S-matrix may be given in the form
  \begin {equation}{{\bf S}={\bf S_o}{\bf S_c}} \quad ,
  \quad{\bf S_o}={S_o}^w \quad ,\quad w=W/\Delta\end {equation} 
  the factor ${S_o}^w$ gives the contribution of the 
  background scattering.
  \par
 When one uses the assumption that $y<<x\;i.e.$ that the spots cover a small
 part of the transverse area it is possible
    to give a simple expression to ${\bf S_c}$
   \begin {equation}{\bf S_c}=1-w y^2 \Delta
  [\rho_b(1-\hat\sigma \rho_f)+ \rho_f(1-\hat\sigma
   \rho_b)] \end {equation}
   It may be noticed that the factor, $w y^2 \Delta$, represents the part
   covered by black spots within the interacting area of the hadrons at given
   {\bf B}.
  
 \vskip 1pc
 The basic elements for calculating the pair and double pair production have
 already been given.
 The local production amplitude is the sum of 4 terms, it will be
 indicated as: 
 $<X_j>=[x^4 X_{o,o}+x^2 y^2 (X_{s,o}+X_{o,s})+y^4 X_{s,s}]\;.$
 The first term represents the pure background interaction, the second and third
 terms
 represent the spot background interaction, the fourth term gives the spot-spot
  interaction. 
  When one consider explicitly the superposition of the two disks it appears
   that the four terms have the same geometrical properties and we get 
  for $\sigma_{\rm eff}$ the same expression as in the case of a uniform hadron.
   So, at first sight it seems that nothing is gained by introducing an
  inhomogeneity into the hadron, but in fact some new features are present.
  The expression of $\sigma_{\rm eff}$ is purely geometrical, it does not
  contain the parameters $x,y$. The expression of $\sigma_{in}$ can be obtained
  from the S-matrix (see eq.28, 29), so it depends on $y$, more in general both on
  $x$ and on $y$.
 \par
  In order to give a clear, although unrealistic evidence of this fact
  we can consider the limit in which the background is so thin that it
  contribute negligibly in the inelastic cross section, then the S-matrix
  element, depends only on the spot-spot interaction 
  In this situation the inelastic
  cross section goes to zero with $y^4$, the corresponding picture would be: few,
  wholly black spots distributed in a wide and very thin background.
   The conclusion of this analysis is therefore that in order to have
  $\sigma_{\rm eff}<\sigma_{in}$ the hadron should be compact, $i.e.$ without
  holes or transparent regions and also with quite sharp edges.
 \vskip 1pc
 \section{ Further considerations and conclusions}
 \vskip 1pc
 The model allows further derivations that are here only mentioned: it is
 possible to introduce correlations in {\bf b} in the case of nonuniform hadron;
 it is also possible to treat to some extent the longitudinal degrees of freedom
 provided the distinction between forward and backward particles is kept valid.
 Moreover, in some simple cases analytically and in more general situation in
 numerical way, it is possible to go beyond the two extreme cases that have
 been illustated here $i.e.$ the weakly interacting or the totally absorbing
 situation.
 \par
 The model presented allows a systematization of
  different aspects of the hard processes in multiparticle production with a
  particular attention to the unitarity corrections and
  suggests also some interpretations in terms of hadron structure.
  The connection with QCD is not direct as it appears from the fact that the
  interaction term is quartic while the fundamental QCD interaction term is
  cubic, in other words the fundamental input is the parton hard scattering, 
  not the branching process.

\section*{Acknowledgments}
This work has been partially supported by the Italian Ministery of University
 and of Scientific and Technological Research by means of the {\it Fondi per la
 Ricerca scientifica - Universit\`a di Trieste }.
\vfill
\eject

\vfill


\section*{References}
\begin{thebibliography}{99}
\bibitem{1} R.J. Glauber, in {\it Lectures in Theoretical Physics} 
 ed. W.E. Brittin (New York, 1959).
 M.M.Islam, in {\it Lectures in Theoretical Physics} vol 9B
 (New York, 1967).
\bibitem{2} Ll.Ametller, D.Treleani {\em Int. J. Mod. Phys.} {\bf A3}, 521
(1988).  
        G. Calucci, D. Treleani {\em Phys. Rev.} {\bf D 41},3367 (1990).
        G.Calucci, D.Treleani,{\em  Phys. Rev.} {\bf D 44},2746 (1991)
\bibitem{3} H.M. Fried {\it Functional methods and models in QFT, ch.9 } M.I.T. 
 Press Cambridge, U.S.A. (1973)
\bibitem{4} J.D. Bjorken, in {\it Multiparticle dynamics 1994 } ed. A.Giovannini,
S.Lupia, R.Ugoccioni (Singapore 1995)
\bibitem{5} O. Nachtmann, High-energy collisions and nonperturbative QCD in {\it
  Lectures on QCD - Applications } ed. F.Lenz, H.Grie{\ss}hammer, D.Stoll, 
  (Springer 1997)
\bibitem{6} M. Drees and T. Han, {\em Phys. Rev. Lett.} {\bf 77}, 4142 (1996).
 F. Abe et al., (CDF Collaboration),{\em Phys. Rev.} {\bf D 56}, 3811 (1997)
\bibitem{7} G.Calucci, D.Treleani {\em Nucl.Phys. B(Proc. Suppl.)} {\bf 71},392 (1999)

\end{thebibliography}
\end{document}